\documentclass{article}
\usepackage{arxiv}
\usepackage[utf8]{inputenc}
\usepackage[T1]{fontenc}
\usepackage{hyperref}
\usepackage{url}
\usepackage{booktabs}
\usepackage{amsfonts}
\usepackage{amsmath}
\usepackage{amssymb}
\usepackage{microtype}
\usepackage{graphicx}
\usepackage[numbers]{natbib}
\usepackage{doi}

\title{Chirp-controlled plasma wake excitation by an exponential laser pulse in underdense plasma}

\author{Ajit Kumar Kushwaha$^1$, Dinkar Mishra$^1$\orcid{0000-0002-7322-5468}, Shivani Aggarwal$^1$\orcid{0009-0005-1537-7374}, Saumya Singh$^1$\orcid{0000-0003-4300-8910}, and Bhupesh Kumar$^{1,*}$\orcid{0000-0002-3926-0657}}

\affil{$^1$Department of Physics, University of Lucknow, Uttar Pradesh, India - 226007}
\affil{$^*$Author to whom any correspondence should be addressed.}
\email{bhupeshk05@gmail.com}

\keywords{
Plasma wakefield acceleration,
Chirped laser pulses,
Exponential frequency chirp,
Relativistic fluid model,
Particle-in-cell simulation,
Laser--plasma interaction,
Electron acceleration
}

\begin{document}
\maketitle

\begin{abstract}
The excitation of plasma wakefields driven by chirped laser pulses is investigated using a reduced relativistic fluid--Poisson model supported by fully relativistic particle-in-cell (PIC) simulations. The study considers exponential, linear, quadratic, and unchirped phase-modulated laser drivers propagating in an underdense plasma. Numerical solutions of the governing equations demonstrate that exponential chirping produces enhanced wakefield amplitudes compared to polynomial and unchirped cases due to nonlinear phase variation across the pulse envelope. The analytical predictions are validated using quasi-cylindrical PIC simulations performed under identical plasma and laser parameters. The simulations reveal strong chirp-dependent wakefield modification, with positively chirped pulses generating peak accelerating fields exceeding $58~\mathrm{GV/m}$, accompanied by pronounced density compression and enhanced electron momentum gain. These results demonstrate that exponential chirping provides an effective mechanism for controlling wakefield strength and improving plasma-based particle acceleration.
\end{abstract}

\section{Introduction}
The interaction of an intense laser pulse with underdense plasma gives rise to several nonlinear phenomena, including wakefield generation [1], self-modulation [2,3], self-focusing [4--7], Raman scattering, and other parametric instabilities [8--11]. These processes are central to many applications such as terahertz radiation generation [12--15], inertial confinement fusion [16], charged-particle acceleration [1,17,18], and harmonic generation [19--23].

Laser--plasma interaction is governed by both laser and plasma parameters. As the laser pulse propagates, plasma electrons respond to the ponderomotive force, which drives an electron plasma wave whose phase velocity is close to the speed of light [24]. For a short pulse, this force displaces electrons from the high-intensity region and generates a large-amplitude electrostatic wakefield [17]. The seminal proposal by Tajima and Dawson established the laser wakefield accelerator (LWFA) concept, in which an ultrashort, high-intensity pulse excites a wake that can trap and accelerate electrons [1]. The development of chirped-pulse amplification made such ultrashort high-power drivers experimentally accessible [25].

Over the past decades, several laser and plasma configurations have been explored to enhance wakefield amplitude and electron energy gain. Grigoriadis \textit{et al}. reported that, under suitable conditions, positively chirped pulses can improve electron acceleration even at relatively low peak intensities, with chirp effects strongly coupled to plasma dispersion [26]. Afhami and Eslami showed that chirp sign and magnitude significantly influence nonlinear wake excitation, with positive and negative chirps producing opposite trends in wake amplitude [27]. Singh \textit{et al}. demonstrated optimization of wakefield generation and acceleration for positively chirped pulses in plasma channels, identifying a critical chirp for maximum wake amplitude, reduced injection energy, and improved energy gain [28]. Ghotra also reported enhanced acceleration with chirped circularly polarized drivers in plasma channels [29]. Related studies have compared chirped and unchirped pulses for different pulse shapes and durations [30].

The present work is motivated by wakefield excitation behind exponentially chirped Gaussian laser pulses. Exponential chirping offers an additional control parameter through temporal phase engineering, which can modify pulse evolution and wake excitation in plasma. Here, we present an analytical and simulation-based study for linearly polarized Gaussian pulses propagating in homogeneous plasma. The analytical model is developed using a perturbative framework together with the quasi-static approximation (QSA). The paper is organized as follows: Sec.~II presents the mathematical formulation based on fluid equations for laser--plasma interaction; Sec.~III provides graphical analysis; Sec.~IV discusses analytical and simulation results for wakefield generation; and Sec.~V summarizes the main conclusions.

\section{Numerical Results}
We consider the propagation of an intense linearly polarized laser pulse along the $z$-direction in a homogeneous underdense plasma with unperturbed electron density $n_0$, working in the regime $\omega_0 \gg \omega_p$ so that the laser dispersion relation reduces to its vacuum form $k \simeq \omega/c$. The ions are assumed to form a stationary neutralizing background over the characteristic time scale of wake excitation, which is justified under the immobile-ion approximation. The laser field is described using a prescribed vector potential profile, while the plasma response is modeled using a relativistic cold electron-fluid formulation closed by the continuity equation and Poisson's equation.

To investigate the influence of chirped laser pulses on wakefield generation, the transverse vector potential of the laser is written as

\begin{equation}
\vec{A}(z,t)=\hat{y}\,A_0\,f(\xi)\cos[\psi(\xi)],
\label{eq1}
\end{equation}

where $\xi = z - ct$ denotes the co-moving coordinate, $A_0$ is the peak vector potential amplitude, and $\hat{y}$ represents linear polarization along the transverse direction. The slowly varying envelope is taken to be Gaussian, $f(\xi) = \exp(-\xi^{2}/2L^{2})$, where $L$ characterizes the pulse length. The phase $\psi(\xi)$ describes the chirped carrier oscillation and is responsible for introducing frequency modulation along the propagation direction. The slowly varying envelope approximation (SVEA) is assumed throughout, requiring the carrier phase to vary much faster than the envelope, $|d\psi/d\xi| \gg |d\ln f/d\xi|$.

In the present work, primary attention is given to exponentially chirped laser pulses. The instantaneous frequency of the laser is prescribed as
\begin{equation}
\omega(\xi) = \omega_0\, e^{-b\xi},
\label{eq:omega}
\end{equation}
where $\omega_0$ denotes the reference frequency and $b$ is the exponential chirp parameter governing the rate of frequency variation along the pulse. The instantaneous frequency in the co-moving frame is related to the phase by $\omega(\xi) = c\,d\psi/d\xi$, so that the phase function is obtained by direct integration,

\begin{equation}
\psi(\xi)=\frac{1}{c}\int \omega(\xi)\,d\xi=\frac{\omega_0}{c}\int e^{-b\xi}\,d\xi.
\label{eq2}
\end{equation}

The integration constant is omitted: under SVEA the wake dynamics depend on the envelope-averaged ponderomotive force, which is insensitive to a uniform carrier-envelope phase shift. Performing the integration yields

\begin{equation}
\psi(\xi)=-\frac{\omega_0}{bc}\,e^{-b\xi}.
\label{eq3}
\end{equation}

The exponential chirp formulation is particularly advantageous because it provides a unified representation from which unchirped, linearly chirped, and quadratically chirped carriers can be recovered through Taylor-series expansion. For $|b\xi| \ll 1$ (equivalently $bL \ll 1$, which is also the regime in which SVEA remains valid for this profile),

\begin{equation}
e^{-b\xi} \approx 1 - b\xi + \frac{1}{2}b^{2}\xi^{2} - \frac{1}{6}b^{3}\xi^{3},
\end{equation}

so that, dropping the irrelevant additive constant, the phase takes the polynomial form

\begin{equation}
\psi(\xi) \approx k_0\,\xi - \frac{1}{2}k_0\,b\,\xi^{2} + \frac{1}{6}k_0\,b^{2}\,\xi^{3}, \qquad k_0 \equiv \frac{\omega_0}{c}.
\label{eq:expansion}
\end{equation}

The corresponding instantaneous frequency, $\omega(\xi) = c\,d\psi/d\xi \approx \omega_0\bigl(1 - b\xi + \frac{1}{2}b^{2}\xi^{2}\bigr)$, makes the chirp hierarchy explicit: retaining the phase up to $\xi$ alone gives an unchirped carrier of frequency $\omega_0$; retaining the $\xi^{2}$ term yields a linear chirp (instantaneous frequency varying linearly with $\xi$); and retaining the $\xi^{3}$ term yields a quadratic chirp (instantaneous frequency varying quadratically with $\xi$). The sign of the $\xi^{2}$ coefficient, set by $b$, determines whether the pulse is down-chirped ($b>0$) or up-chirped ($b<0$). This hierarchical structure allows direct comparison between exponential, linear, and quadratic chirped drivers within the same computational framework.

\begin{figure}[t]
\centering
\includegraphics[width=0.9\linewidth]{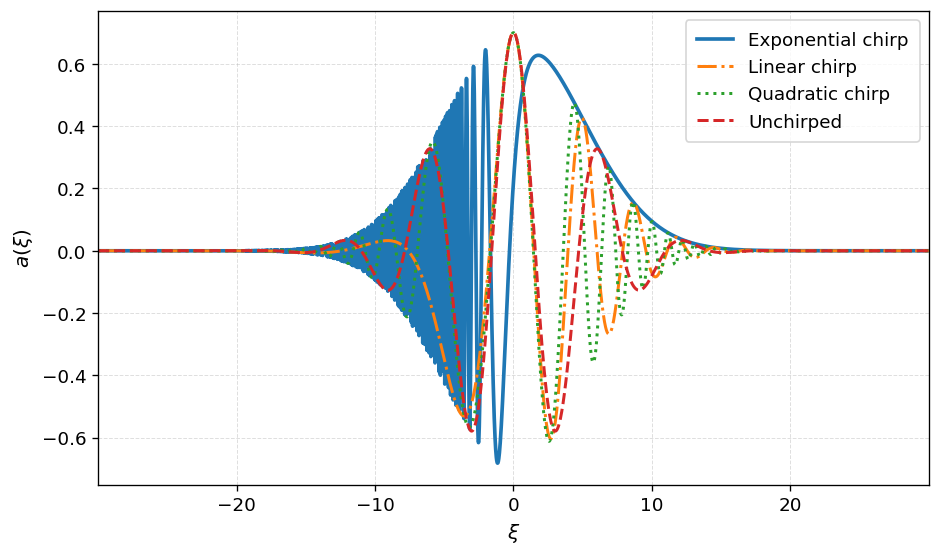}
\caption{Normalized laser field profiles $a(\xi)$ for exponential, linear, quadratic, and unchirped phase modulations. The exponential chirp produces strong nonlinear phase distortion across the pulse envelope, whereas linear and quadratic chirps introduce progressively weaker frequency modulation. The unchirped pulse retains symmetric oscillations and serves as a reference case.}
\label{fig:1}
\end{figure}

To illustrate the influence of chirping on the laser driver, the normalized vector potential profiles corresponding to exponential, linear, quadratic, and unchirped phase modulations are shown in Fig.~\ref{fig:1}. The exponential chirp produces a strongly nonlinear phase variation across the pulse envelope, resulting in asymmetric oscillation spacing. In contrast, linear and quadratic chirps introduce progressively weaker phase distortion, while the unchirped case retains symmetric oscillations. These profiles serve as input drivers for the subsequent wakefield calculations.

The plasma response to the applied laser field is described using the relativistic cold electron-fluid equations,

\begin{equation}
\frac{d(\gamma\vec{v})}{dt}=-\frac{e}{m}\left[-\nabla\varphi-\frac{1}{c}\frac{\partial\vec{A}}{\partial t}+\frac{\vec{v}}{c}\times(\nabla\times\vec{A})\right],
\label{eq4}
\end{equation}

\begin{equation}
\frac{\partial n_e}{\partial t}+\nabla\cdot(n_e\vec{v})=0,
\label{eq5}
\end{equation}

\begin{equation}
\nabla^2\varphi=-4\pi e(n_0-n_e).
\label{eq6}
\end{equation}

Here $\vec{v}$ denotes the electron-fluid velocity, $n_e$ is the electron density, $\varphi$ is the electrostatic potential, and $\gamma=(1-v^2/c^2)^{-1/2}$ is the relativistic Lorentz factor. These equations collectively describe momentum conservation, charge continuity, and electrostatic field generation in the plasma medium.

To separate the rapid oscillatory motion associated with the laser field from the slowly varying plasma response, the electron velocity is decomposed into fast and slow components, $\vec{v}=\vec{v}_f+\vec{v}_s$. For a linearly polarized configuration with polarization along $y$ and propagation along $z$, the dominant fast motion occurs in the transverse direction. The fast quiver velocity follows directly from the transverse component of equation~(\ref{eq4}), leading to $v_{yf}=(ac/\gamma)$, where $a=eA/(mc^2)$ is the normalized vector potential amplitude. The remaining components of the fast velocity vanish in the one-dimensional approximation.

Substituting the velocity decomposition into the governing equations and averaging over the fast oscillation period yields the slow-time-scale dynamics that govern wakefield formation. Using the convective derivative $d/dt=\partial/\partial t+v_z\partial/\partial z$, and adopting the quasi-static approximation appropriate for ultrarelativistic laser pulses, the temporal derivatives may be replaced by spatial derivatives in the co-moving frame.

Introducing normalized quantities $u=v/c$, $n=n_e/n_0$, and $\phi=e\varphi/(mc^2)$, and defining the plasma wavenumber $k_p=\omega_p/c$ with $\omega_p=(4\pi n_0 e^2/m)^{1/2}$, the resulting normalized slow-scale equations describing the plasma response can be written as

\begin{equation}
\frac{\partial u_{ys}}{\partial(k_p\xi)}=\frac{1}{\gamma}\frac{\partial a}{\partial(k_p\xi)}-\frac{u_{ys}}{\gamma}\frac{\partial\gamma}{\partial(k_p\xi)},
\label{eq7}
\end{equation}

\begin{equation}
\frac{\partial u_{zs}}{\partial(k_p\xi)}=\frac{1}{\gamma(1-u_{zs})}\left[\frac{\partial\phi}{\partial(k_p\xi)}-u_{ys}\frac{\partial a}{\partial(k_p\xi)}\right]-\frac{u_{zs}}{\gamma}\frac{\partial\gamma}{\partial(k_p\xi)},
\label{eq8}
\end{equation}

\begin{equation}
\frac{\partial n}{\partial(k_p\xi)}=\frac{n}{1-u_{zs}}\frac{\partial u_{zs}}{\partial(k_p\xi)},
\label{eq9}
\end{equation}

\begin{equation}
\frac{\partial^2\phi}{\partial(k_p\xi)^2}=n-1.
\label{eq10}
\end{equation}

These equations represent, respectively, the transverse electron response driven by the laser field, longitudinal momentum evolution governed by ponderomotive coupling, density perturbation growth, and electrostatic restoring forces responsible for wakefield formation.

The normalized vector potential entering these equations is expressed as

\begin{equation}
a(\xi)=a_0\exp\left(-\frac{\xi^2}{2L^2}\right)\cos[\psi(\xi)],
\label{eq11}
\end{equation}

where $a_0=eA_0/(mc^2)$ is the normalized peak amplitude. The derivative $\partial a/\partial\xi$ appearing in equations~(\ref{eq7}) and (\ref{eq8}) incorporates both envelope and phase variation, thereby allowing the chirped frequency structure to influence the wakefield excitation process through the ponderomotive force.

The electrostatic potential obtained from equation~(\ref{eq10}) determines the longitudinal accelerating field, which is calculated from

\begin{equation}
E_z=-\frac{\partial\phi}{\partial\xi}.
\label{eq12}
\end{equation}

The longitudinal electric field represents the primary accelerating component of the plasma wake and determines the energy gain of charged particles interacting with the wakefield structure.

The coupled nonlinear equations (\ref{eq7})--(\ref{eq10}) are solved numerically using a fourth-order Runge--Kutta integration scheme along the co-moving coordinate $\xi$. The integration is performed for prescribed values of the normalized laser amplitude $a_0$, pulse length $L$, chirp parameter $b$, and plasma density $n_0$, corresponding to the laser and plasma parameters used in the subsequent particle-in-cell simulations.

\begin{figure}[t]
\centering
\includegraphics[width=0.9\linewidth]{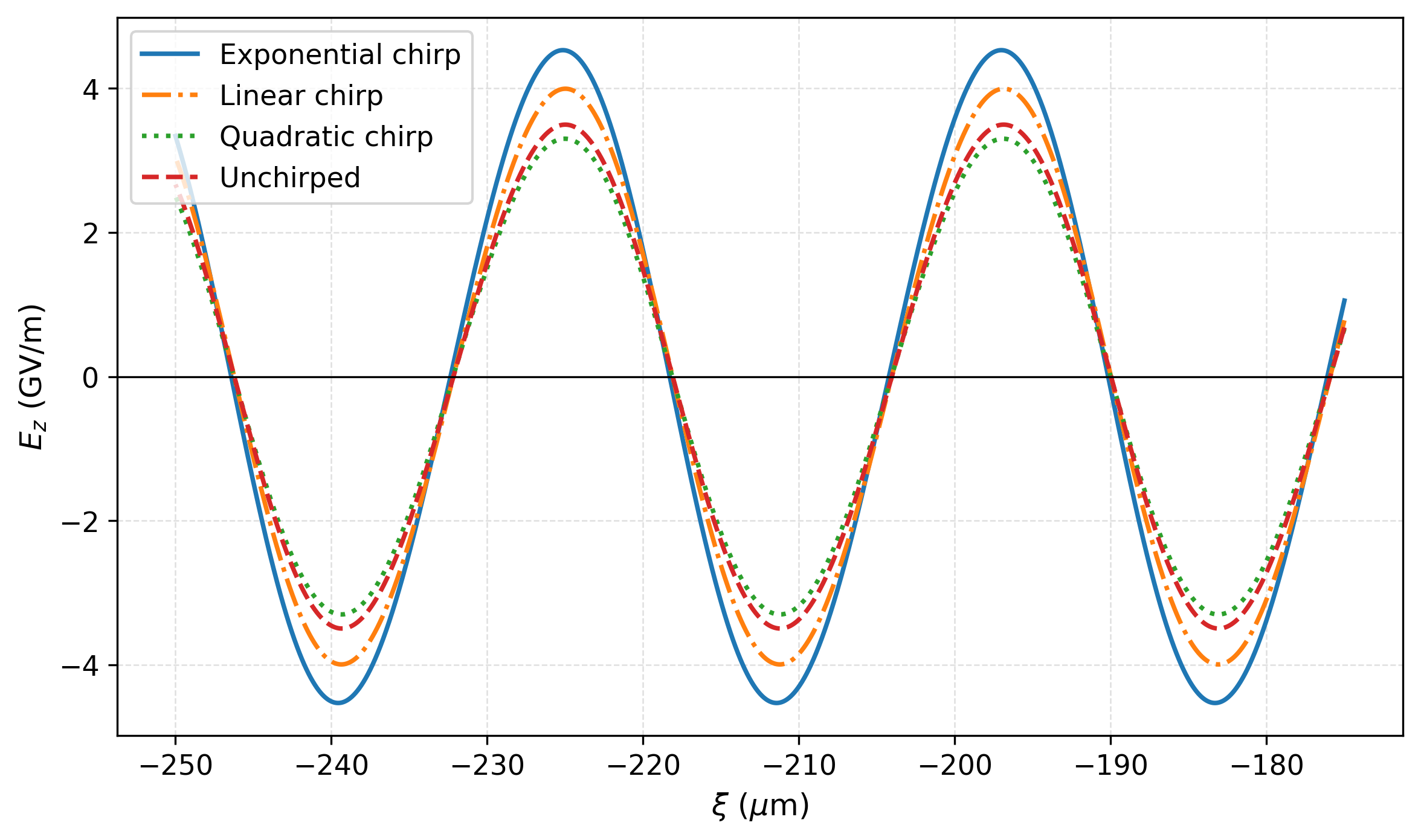}
\caption{Longitudinal wakefield profiles $E_z(\xi)$ obtained from numerical integration of the reduced fluid--Poisson model for exponential, linear, quadratic, and unchirped laser pulses. The exponential chirp produces a pronounced modification of the wake structure due to nonlinear phase variation across the pulse envelope, while polynomial chirps introduce progressively weaker distortions relative to the unchirped reference case. These analytical wakefield profiles serve as baseline predictions for comparison with particle-in-cell simulation results presented in the subsequent section.}
\label{fig:2}
\end{figure}

The primary quantity of interest obtained from the numerical solution is the longitudinal wakefield $E_z(\xi)$, which is calculated from the electrostatic potential through equation~(\ref{eq12}). This field represents the accelerating structure driven by the chirped laser pulse and serves as the principal observable for comparison between different chirp configurations. Representative wakefield profiles obtained for exponential, linear, quadratic, and unchirped laser drivers are presented in Fig.~\ref{fig:2}.

Figure~\ref{fig:2} shows the longitudinal wakefield profiles $E_z(\xi)$ obtained from numerical integration of equations~(\ref{eq7})--(\ref{eq10}) using the chirped driver defined by equation~(\ref{eq11}) and the wakefield relation given by equation~(\ref{eq12}). The calculations were performed for laser wavelength $\lambda_0=0.8~\mu$m, plasma density $n_0=1.41\times10^{18}~\mathrm{cm^{-3}}$, normalized laser amplitude $a_0=0.7$, and pulse duration $\tau=5~\mathrm{fs}$. The chirp parameters were taken as $b=0.8$ for the exponential case, $b_1=0.3$ for the linear case, and $(b_1,b_2)=(0.3,0.05)$ for the quadratic case. The values of the polynomial chirp parameters $b_1$ and $b_2$ were selected such that the linear and quadratic phase variations represent the leading-order Taylor expansion terms of the exponential chirp under small $b\xi$ conditions.

A clear dependence of the wake amplitude on the chirp profile is observed. The exponential chirp produces the largest peak accelerating field of approximately $4.55~\mathrm{GV/m}$, followed by the linear chirp with a maximum amplitude of about $4.00~\mathrm{GV/m}$. The unchirped pulse yields a peak field of approximately $3.55~\mathrm{GV/m}$, while the quadratic chirp results in the lowest peak amplitude of about $3.35~\mathrm{GV/m}$. The enhanced wakefield obtained for the exponential chirp indicates more efficient ponderomotive driving due to the nonlinear variation of the instantaneous phase, whereas polynomial chirps introduce comparatively weaker modifications to the wake structure under the present parameter conditions. For comparison with particle-in-cell simulations, the co-moving coordinate $\xi$ used in the analytical calculations corresponds directly to the spatial propagation coordinate $z$ in the moving simulation window.

\begin{figure}[t]
\centering
\includegraphics[width=0.9\linewidth]{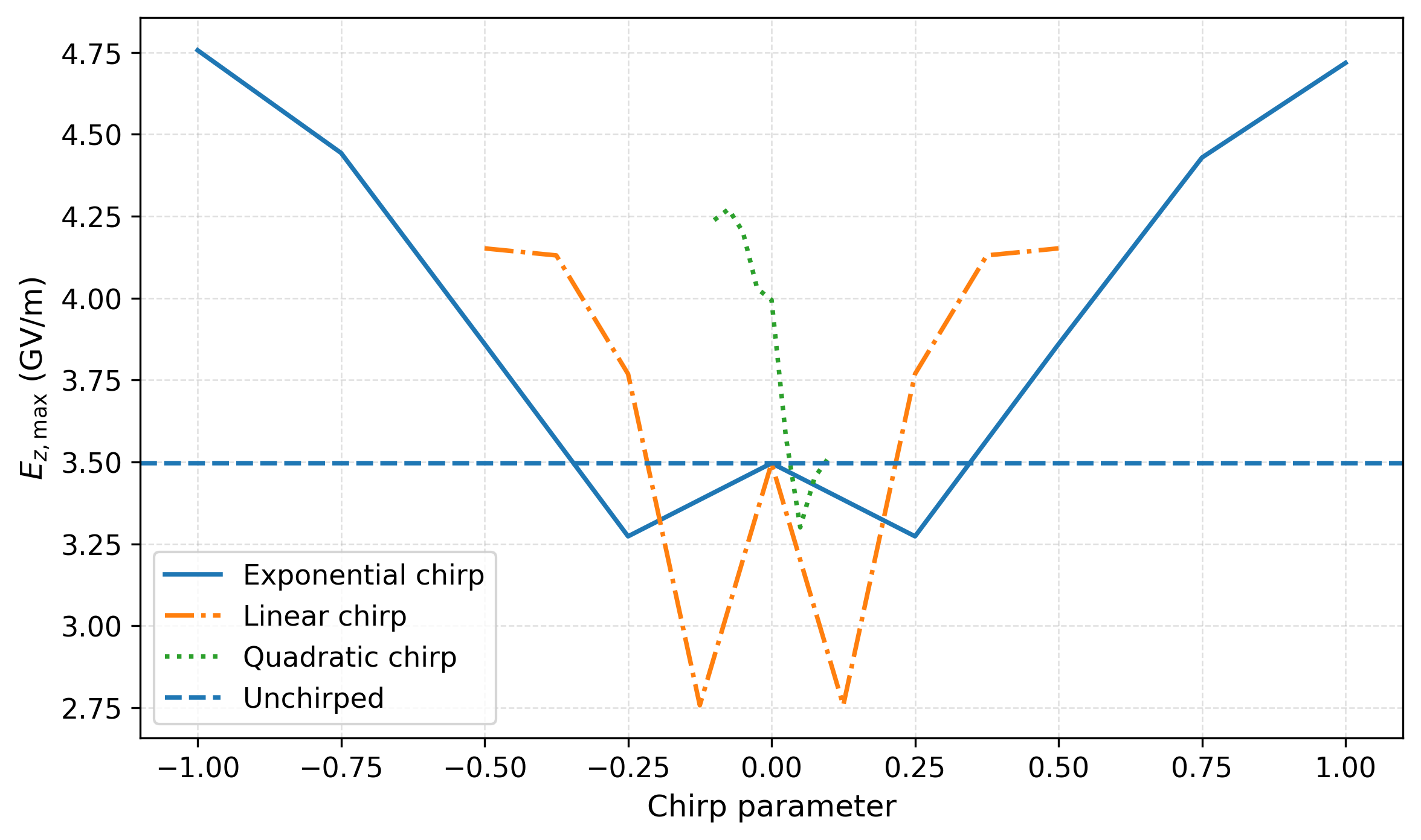}
\caption{Maximum longitudinal wakefield amplitude $E_{z,\max}$ as a function of chirp parameter for exponential, linear, and quadratic chirped laser pulses. The horizontal dashed line represents the unchirped reference case. Both positive and negative exponential chirps produce significant enhancement of the wakefield, while linear and quadratic chirps exhibit non-monotonic behavior with regions of suppressed and enhanced wake excitation.}
\label{fig:3}
\end{figure}

Figure~\ref{fig:3} shows the variation of the maximum longitudinal wakefield amplitude $E_{z,\max}$ as a function of chirp parameter for exponential, linear, and quadratic chirped laser pulses obtained from numerical integration of equations~(\ref{eq7})--(\ref{eq10}). The unchirped reference level corresponds to a peak field of approximately $3.55~\mathrm{GV/m}$, shown as the horizontal dashed line.

A pronounced dependence of the wake amplitude on the chirp sign and magnitude is observed. For the exponential chirp, both positive and negative chirps enhance the wakefield relative to the unchirped case, with the maximum values reaching approximately $4.75~\mathrm{GV/m}$ at $b=-1.0$ and $4.72~\mathrm{GV/m}$ at $b=1.0$. The wake amplitude decreases toward the unchirped value as the chirp parameter approaches zero, reaching a minimum of approximately $3.28~\mathrm{GV/m}$ near $|b|\approx0.25$.

In contrast, the linear chirp exhibits a non-monotonic behavior in which moderate negative chirp ($b_1\approx-0.1$) produces the lowest peak field of approximately $2.75~\mathrm{GV/m}$, indicating a suppression of wake excitation relative to the unchirped case. Positive linear chirp gradually recovers and enhances the wake amplitude, reaching values near $4.15~\mathrm{GV/m}$ for $b_1\approx0.5$.

The quadratic chirp shows comparatively weaker sensitivity to the chirp parameter, with peak fields remaining within the range $3.30$--$4.25~\mathrm{GV/m}$ for the values considered. A slight asymmetry between positive and negative quadratic chirp is observed, with negative values producing marginally larger wake amplitudes.

The overall trends indicate that negative chirp generally produces stronger wake excitation than positive chirp under the present parameter conditions. This behavior arises from the chirp-induced modification of the local phase velocity and ponderomotive force distribution, which alters the efficiency of energy transfer from the laser pulse to the plasma wake. These results further confirm the superior effectiveness of exponential chirping for enhancing wakefield amplitude compared to polynomial chirp models. The approximately symmetric dependence of the wakefield amplitude on positive and negative exponential chirp arises from the nonlinear nature of the exponential phase modulation, which modifies the local ponderomotive force distribution in a comparable manner for opposite chirp signs under the present parameter regime.

The optimal chirp conditions corresponding to the maximum wakefield amplitude were obtained from the data shown in Fig.~\ref{fig:3}. For the unchirped pulse, the peak accelerating field is approximately $3.55~\mathrm{GV/m}$. The exponential chirp exhibits the strongest enhancement, reaching a maximum field of approximately $4.75~\mathrm{GV/m}$ at $b=-1.0$, corresponding to an increase of about $34\%$ relative to the unchirped case.

For the linear chirp, the maximum field of approximately $4.15~\mathrm{GV/m}$ occurs near $b_1\approx0.5$, representing an enhancement of about $17\%$. The quadratic chirp produces a peak field of approximately $4.25~\mathrm{GV/m}$ near $b_2\approx-0.05$, corresponding to an enhancement of approximately $20\%$. These results indicate that exponential chirping provides the most effective mechanism for increasing wakefield amplitude under the present plasma and laser conditions. The stronger enhancement observed for negative exponential chirp suggests improved phase matching between the laser driver and plasma oscillation, resulting in more efficient energy transfer to the wakefield.

\section{Simulation Results}
To validate the analytical predictions obtained from the reduced fluid--Poisson model, particle-in-cell (PIC) simulations were performed using the fully relativistic quasi-cylindrical code FBPIC. The simulations were carried out in cylindrical geometry with two azimuthal modes ($N_m=2$), which is sufficient to accurately capture the dominant axisymmetric wakefield dynamics while maintaining computational efficiency.

The computational domain extended from $z_{\min}=-10~\mu$m to $z_{\max}=120~\mu$m with $N_z=1800$ grid points along the propagation direction and $N_r=150$ grid points in the radial direction up to $r_{\max}=25~\mu$m. A moving window propagating at the speed of light ($v=c$) was employed to follow the laser pulse over the interaction length. Open boundary conditions were applied in both the longitudinal and radial directions.

The plasma density was set to $n_0=1.41\times10^{18}~\mathrm{cm^{-3}}$, consistent with the parameters used in the analytical model. The plasma column extended up to $z = 300~\mu$m along the propagation direction. A smooth density ramp of length $10~\mu$m was introduced at the plasma entrance and exit to minimize numerical reflections and facilitate stable wake formation. Beyond the plasma termination point, the medium was treated as vacuum, and therefore the electrostatic wakefields were not sustained.

The laser pulse was modeled using a Gaussian transverse profile with waist $w_0=20~\mu$m and a chirped longitudinal profile corresponding to the analytical driver defined in equation~(\ref{eq11}). The central wavelength was taken as $\lambda_0=0.8~\mu$m, with normalized amplitude $a_0=0.7$ and pulse duration $\tau=5~\mathrm{fs}$. The exponential chirp parameter was varied in separate simulations to investigate its influence on wakefield generation.

Field and particle diagnostics were recorded at regular intervals during the simulations. The longitudinal electric field component $E_z$, electron density perturbation $n_e/n_0$, and electron phase-space distributions were extracted along the propagation axis to provide direct comparison with the analytical predictions.

\begin{figure}[t]
\centering
\includegraphics[width=0.9\linewidth]{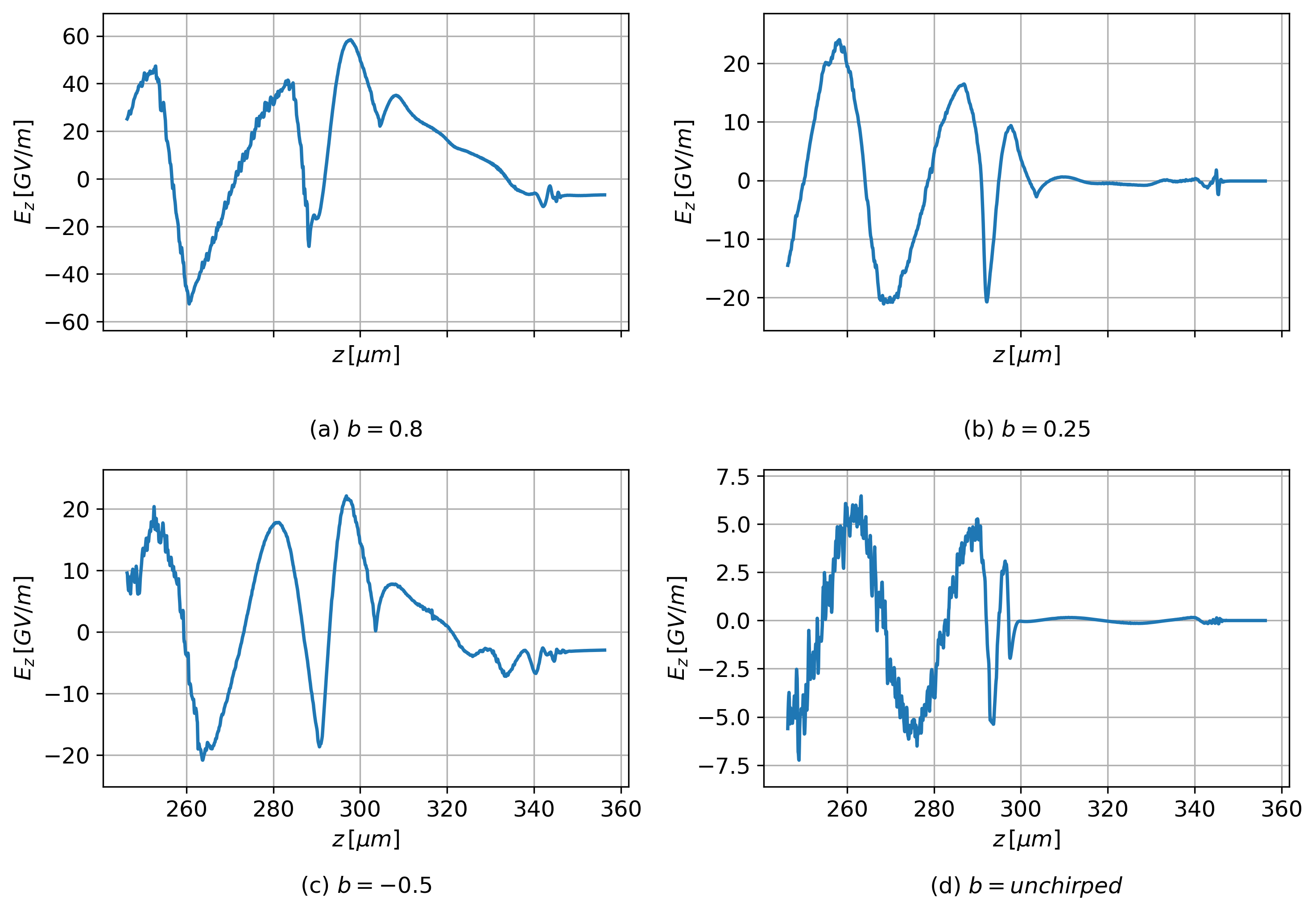}
\caption{Longitudinal wakefield profiles $E_z$ obtained from particle-in-cell simulations for different values of the exponential chirp parameter: (a) $b=0.8$, (b) $b=0.25$, (c) $b=-0.5$, and (d) unchirped pulse ($b=0$). The wakefields are plotted along the propagation coordinate $z$ after excluding the laser region to isolate the plasma wake structure.}
\label{fig:wakefield_profiles}
\end{figure}

The longitudinal wakefield profiles obtained from PIC simulations are presented in Fig.~\ref{fig:wakefield_profiles} for representative values of the chirp parameter. The longitudinal electric field $E_z$ is plotted along the propagation coordinate $z$ at the final simulation time, after excluding the laser region to isolate the wakefield structure.

A strong dependence of the wake amplitude on the chirp parameter is clearly observed. In particular, the positively chirped pulse with $b=0.8$ generates the strongest wakefield, reaching peak amplitudes of approximately $58~\mathrm{GV/m}$, while the unchirped pulse produces significantly weaker fields with peak values near $7~\mathrm{GV/m}$. Intermediate chirp values produce moderate wake amplitudes, consistent with the trends predicted by the analytical model.

In addition to amplitude variation, the spatial structure of the wakefield also exhibits chirp-dependent modulation. Strong positive chirp leads to enhanced plasma oscillations and sharper field gradients, whereas weaker or negative chirp values produce comparatively smoother wake structures. These results demonstrate that laser chirping plays a crucial role in controlling the strength and morphology of the plasma wakefield. The disappearance of wake oscillations beyond $z \approx 300~\mu$m corresponds to the termination of the plasma column, beyond which electrostatic wakefields cannot be sustained.

\begin{figure}[t]
\centering
\includegraphics[width=0.9\linewidth]{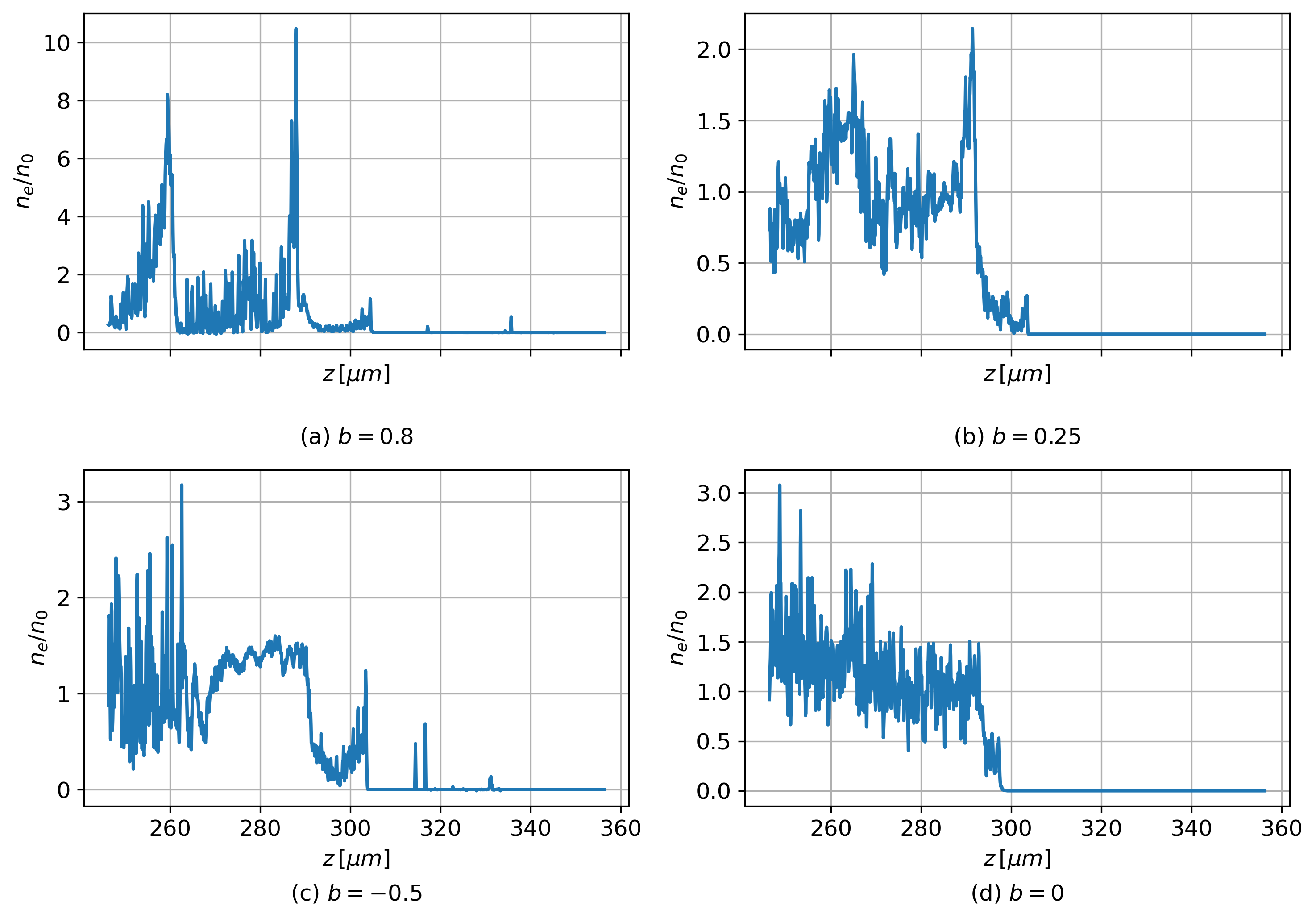}
\caption{Normalized electron density perturbation profiles $n_e/n_0$ obtained from PIC simulations for different values of the exponential chirp parameter: (a) $b=0.8$, (b) $b=0.25$, (c) $b=-0.5$, and (d) unchirped pulse ($b=0$).}
\label{fig:density_profiles}
\end{figure}

The corresponding plasma density perturbations obtained from PIC simulations are shown in Fig.~\ref{fig:density_profiles}. The normalized electron density $n_e/n_0$ exhibits strong modulation behind the laser pulse, forming characteristic density spikes associated with nonlinear plasma oscillations.

A pronounced dependence on the chirp parameter is observed. The positively chirped pulse with $b=0.8$ produces the strongest density compression, with peak density enhancements significantly exceeding those obtained for weaker chirp values and the unchirped case. This enhanced density localization indicates more efficient energy transfer from the laser to the plasma, resulting in stronger wakefield generation as observed in Fig.~\ref{fig:wakefield_profiles}.

For moderate positive chirp ($b=0.25$) and negative chirp ($b=-0.5$), the density oscillations remain present but exhibit reduced peak compression and smoother spatial modulation. In contrast, the unchirped case ($b=0$) shows comparatively weaker density perturbations, consistent with the lower wakefield amplitudes obtained in the corresponding field profiles.

\begin{figure}[t]
\centering
\includegraphics[width=0.9\linewidth]{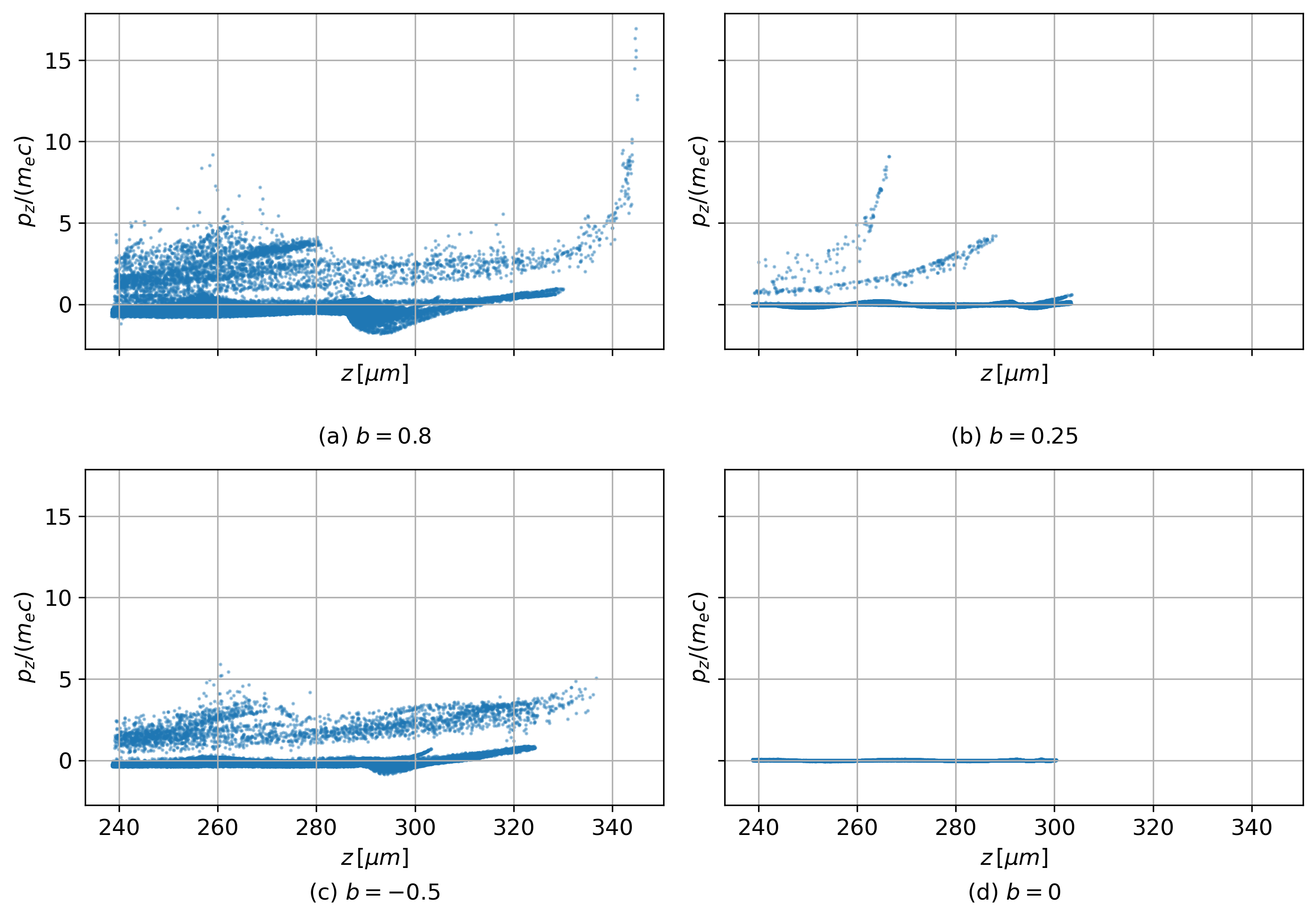}
\caption{Electron phase-space distributions in the $(z, p_z/m_ec)$ plane obtained from PIC simulations for different values of the exponential chirp parameter: (a) $b=0.8$, (b) $b=0.25$, (c) $b=-0.5$, and (d) unchirped pulse ($b=0$).}
\label{fig:phase_space}
\end{figure}

The electron phase-space distributions corresponding to different chirp parameters are presented in Fig.~\ref{fig:phase_space}. The longitudinal momentum profiles clearly reflect the strength of the wakefields generated in each case.

For the strongly positively chirped pulse ($b=0.8$), electrons attain the highest momentum, reaching values exceeding $p_z \approx 15\,m_ec$, accompanied by the formation of dense phase-space structures that indicate efficient energy transfer from the wakefield to the plasma electrons. This behavior is consistent with the strong wakefield amplitudes and pronounced density compression observed in Figs.~\ref{fig:wakefield_profiles} and \ref{fig:density_profiles}.

For moderate positive chirp ($b=0.25$), the phase-space distribution exhibits noticeable acceleration with peak momenta approaching $p_z \approx 9\,m_ec$. In contrast, the negatively chirped pulse ($b=-0.5$) produces comparatively weaker acceleration, with peak momenta limited to approximately $p_z \approx 4\,m_ec$. The unchirped pulse ($b=0$) shows minimal phase-space evolution, indicating negligible electron acceleration and confirming the weak wakefield excitation observed in the corresponding field and density profiles.

Overall, the simulation results demonstrate consistent trends across field, density, and phase-space diagnostics, confirming the strong influence of exponential chirping on plasma wake excitation and electron acceleration. The observed enhancement in wakefield strength and electron momentum for positively chirped pulses supports the analytical predictions and highlights the effectiveness of chirp control as a practical mechanism for optimizing laser-driven plasma acceleration.

\section{Conclusions}
In this work, the generation of plasma wakefields driven by chirped laser pulses has been investigated using a relativistic cold-fluid model coupled with Poisson's equation. The analytical formulation developed in the quasi-static framework enabled systematic evaluation of plasma density perturbations and the associated longitudinal accelerating field under different chirp conditions. Numerical integration of the governing equations demonstrated a clear dependence of wakefield strength on the chirp profile, with exponential chirping producing the largest enhancement in wake amplitude compared to linear, quadratic, and unchirped drivers. The nonlinear phase variation associated with exponential chirp modifies the ponderomotive force distribution and improves the efficiency of wake excitation.

The analytical predictions were validated through fully relativistic particle-in-cell simulations carried out using the quasi-cylindrical FBPIC code under identical plasma and laser conditions. The simulation results confirmed strong chirp-dependent wakefield modification, with positively chirped pulses producing peak accelerating fields of approximately $58~\mathrm{GV/m}$, significantly larger than those obtained for the unchirped case. Corresponding density perturbations exhibited strong nonlinear compression for large chirp values, indicating enhanced plasma response and improved energy coupling. Phase-space analysis further revealed that exponential chirping promotes efficient electron trapping and momentum growth, leading to higher acceleration efficiency compared to weakly chirped or unchirped pulses.

Overall, the combined analytical and simulation results demonstrate that exponential chirping provides a robust and controllable mechanism for enhancing plasma wake excitation and optimizing electron acceleration. The ability to tailor wakefield characteristics through chirp modulation offers promising opportunities for improving compact plasma-based accelerator designs. Future studies may extend the present model to include multidimensional effects, transverse laser evolution, and experimentally realistic plasma configurations to further refine chirp-based wakefield optimization strategies.

\end{document}